\documentclass{article}

\usepackage{subfig}
\usepackage{xspace}
\usepackage{graphicx}
\usepackage[utf8]{inputenc}
\usepackage{soul}
\usepackage{authblk}
\usepackage{lineno}
\usepackage{verbatim}

\newcommand*{\MET}{\ensuremath{E_{\textrm{T}}^{\textrm{miss}}}\xspace}
\newcommand*{\met}{\ensuremath{E_{\textrm{T}}^{\textrm{miss}}}\xspace}

\title{Search for dark matter at colliders}

\author[1]{Oliver Buchmueller}
\affil{High Energy Physics Group, Blackett Laboratory, Imperial College, London, United Kingdom}

\author[2]{Caterina~Doglioni}
\affil[2]{Fysiska institutionen, Lunds universitet, Lund, Sweden}

\author[3]{Lian-Tao~Wang} 
\affil[3]{Corresponding author. Enrico Fermi Institute and Department of Physics, University of Chicago, USA. }

\begin{document}

\maketitle

\begin{abstract}
Multiple astrophysical  and  cosmological  observations  show  that  the majority of the matter in the universe is non-luminous.  It is not made of known particles, and it is called dark matter.  This is one of the few pieces of concrete experimental evidence of  new  physics  beyond  the  Standard Model. Despite decades of effort, we still know very little about the identity of dark matter; it remains one of the biggest outstanding mysteries facing particle physics.  Among the numerous proposals to explain its nature, the Weakly Interacting Massive Particle (WIMP) scenario stands out.  The WIMP scenario is based on a simple assumption that dark matter is in thermal equilibrium in the early hot universe, and that the dark matter particles have mass and interactions not too different from the massive particles in the Standard Model.  Testing the WIMP hypothesis is a focus for many experimental searches.  A variety of techniques are employed including the observation of WIMP annihilation, the measurement of WIMP-nucleon scattering in terrestrial detectors, and the inference of WIMP production at high energy  colliders.  In this article, we will focus on the last approach, and in particular on WIMP dark matter searches at the Large Hadron Collider. \textit{Authors note:} this paper (and references therein) correspond to the version that was submitted to the joint issue of Nature Physics and Nature Astronomy in January 2017. 

\end{abstract}

\section{Introduction}

Several cosmological and astrophysical observations have shown that the vast majority of the matter in the universe is non-luminous. The energy budget of the universe can be categorized as ordinary matter, dark matter (DM) and dark energy~\cite{Bertone:2004pz}. Measurements of the cosmic microwave background, large scale structure and galaxy formation, show that DM comprises 27\% of this energy budget, with ordinary matter only accounting for approximately 5\% \cite{Ade:2015xua} 
Whilst the Standard Model (SM) of particle physics accurately describes the nature and properties of ordinary matter, we know very little about DM. Understanding the underlying nature of DM is one of the biggest unsolved problems in physics. Today, one of the most compelling theoretical explanations for the nature of DM is that it consists of a new class of subatomic particles, known as Weakly Interacting Massive Particles (WIMPs). These particles have masses of the order of the weak scale, corresponding to $\sim$100 gigaelectronvolt (GeV), and only interact weakly with SM particles. 
Such particles would explain the measured amount of DM in the universe~\cite{Gelmini:2015zpa}. Details about this DM relic abundance are given in Box 1.
WIMPs feature in many new physics models, such as Supersymmetry (SUSY), 
which predict new massive particles in order to address several open questions of the Standard Model. The striking concurrence that both cosmology and particle physics independently point to new particles at the weak scale further motivates the WIMP hypothesis and guides a decades-long experimental search programme for DM. 
Even though WIMPs are not the only possible DM scenario, the discovery of WIMPs or the falsification of the WIMP hypothesis will be an essential step forward in the search for DM and hence in our understanding of the Universe.
There is a long road ahead of us in our quest to probe the WIMP dark matter hypothesis. Collider searches for WIMPs, such as the ones performed at the Large Hadron Collider (LHC) at CERN~\cite{Collider:1998498}, are expected to make vital contributions towards this important scientific goal, following a history of searches at the Tevatron~\cite{Electron-Positron:1997351} and LEP~\cite{Tevatron} colliders.

\subsection*{The search for WIMP DM}

\begin{figure}[t!]
	\includegraphics[width=\textwidth]{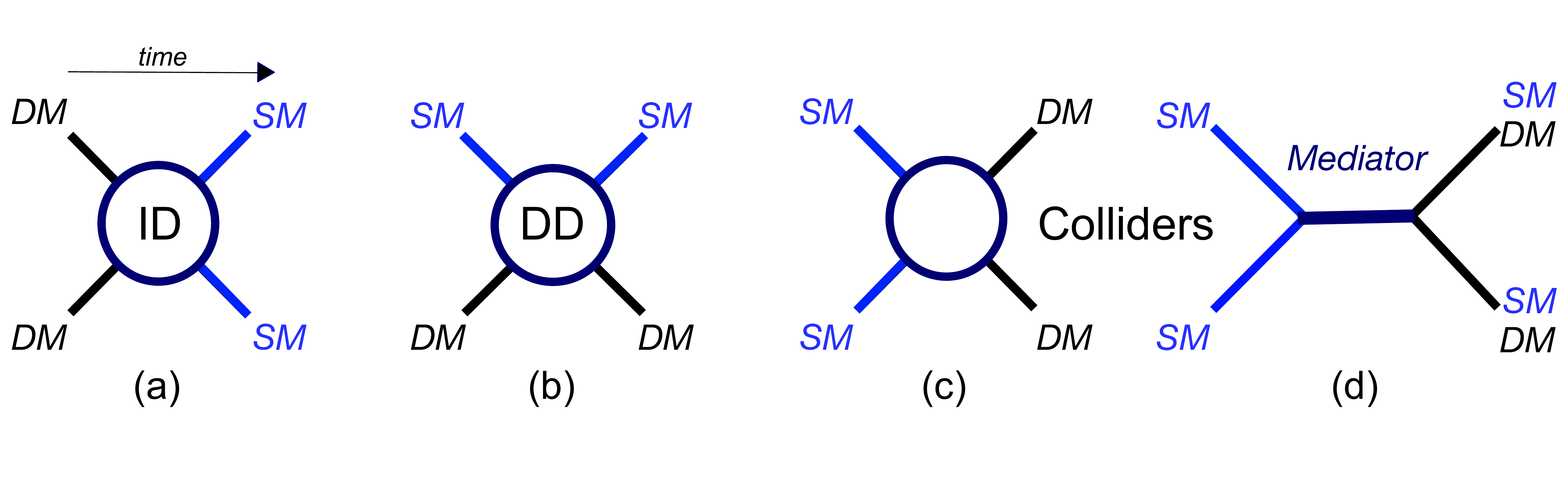}
  \caption{Schematic illustration of DM interactions and their corresponding experimental detection techniques, with time flowing from left to right. Fig. (a) shows DM annihilation to SM particles, sought by Indirect Detection (ID) experiments. Fig. (b) shows DM $\rightarrow$ SM particle scattering, as searched for in Direct Detection (DD) experiments. Fig. (c) shows the production of DM particles from the annihilation of SM particles at colliders. Fig. (d) shows again the pair production of DM at colliders as in (c), but in this specific case the interaction occurs through a mediator particle between DM and SM particles. As demonstrated in the examples shown in last two diagrams, if the theory predicts the creation of DM through some mediator, then the corresponding process with the mediator decaying into SM particles will also occur. Adapted from~\cite{EFTSimplifiedModels}}
    	\label{fig:DMComplementarity}
\end{figure}

The search for WIMP DM is a multi-disciplinary effort that involves different experimental detection techniques. As shown in Fig.~\ref{fig:DMComplementarity} (a-c), DM interactions with SM particles
inspire different detection approaches: 
Indirect Detection (ID) experiments seek detection of DM via decay or annihilation into SM particles (a), Direct Detection (DD) experiment are designed to observe SM-DM scattering (b), and collider experiments aim to detect the DM particles produced in the particle collisions (c-d). Whilst all these experimental search strategies rely on detecting the interactions of DM particles with SM particles, their approaches are orthogonal and complementary. 

ID experiments aim to detect the fluxes of SM products, in particular gamma rays, charged leptons and neutrinos, produced by the annihilation or decay of DM particles~\cite{Gaskins:2016cha}. 
They search for signals of WIMP annihilation from areas with expected higher DM concentration, such as the Galactic Center and dwarf spheroidal galaxies, as well as for signals from WIMPs accumulated in the Sun and Earth. 

DD experiments detect the scattering of galactic DM particles off atomic nuclei by measuring the nuclear recoil energy, which is typically in the kiloelectronvolt range. This is accomplished by sensitive detectors built to minimize background contamination from cosmic, environmental and internal interactions~\cite{Undagoitia:2015gya}. 

Collider searches aim to detect signals from DM particles produced when colliding SM particles (e.g. protons in the case of the LHC), in controlled laboratory conditions.
Collider experiments can search for invisible DM particles, but they can also probe the interaction between the SM and DM particles by searching for the particles that mediate it, as shown in Fig.~\ref{fig:DMComplementarity} (d). 
DD and ID experiments would establish the galactic origin of a stable WIMP signal, with limited sensitivity to the details of the interaction. In a collider experiment, the stability of DM particles is only established up to the timescales needed to traverse the detector. Unlike DD and ID experiments, collider experiments cannot probe whether these particles have a much longer lifetime such as that required for new particles constituting the DM relic density~\cite{0264-9381-25-11-114003}. Therefore, collider searches are only able to ascertain the WIMP stability on the timescale required for these particles to exit the detector, but they can be designed to probe the SM-WIMP interaction in greater detail. 
Compelling evidence of WIMP DM would require confirmation from all these complementary experimental techniques. 

This article focuses on the progress of WIMP DM searches at the LHC, outlining the state of the art of the benchmark models used to design searches, the results, and the future directions for collider searches. 

\section{Collider searches for WIMPs}

 Searches for WIMP dark matter have been carried out at several high energy colliders, such as the Large Electron Positron Collider at CERN and the Tevatron at Fermilab (see e.g. Refs.~\cite{Fox:2011fx,Ellis:207782,Bai:2010hh}). It is the data from the currently operating LHC which provides the greatest sensitivity to rare processes and allows access to the highest energy scales for new phenomena involving Dark Matter. For this reason, this review is focused on WIMP searches at the LHC. 

The LHC is the largest operating particle collider in the world~\cite{Evans:2008zzb}. It is located near Geneva, Switzerland. The four main experiments at the LHC are ATLAS~\cite{PERF-2007-01}, ALICE~\cite{Aamodt:2008zz}, CMS~\cite{CMS-TDR-08-001}, and LHCb~\cite{Alves:2008zz}. The ATLAS and CMS experiments are multi-purpose experiments with similar physics goals: the measurement of Higgs boson properties and for physics beyond the SM, such as the nature of DM, as well as precise measurements of the SM. The ALICE experiment studies heavy ion collisions, while the LHCb experiment is a precision beauty and charm quark physics experiment. 

During its first data-taking period from March 2010 to February 2013, called~\textit{Run 1}, the LHC has delivered proton-proton collisions at a centre-of-mass energies of 7 and 8 teraelectronvolt (TeV). The LHC has restarted operations in summer 2015 (\textit{Run 2}), at a centre-of-mass energies of 13 TeV. Both the LHC collision energy and the ability to deliver and collect collision data have been increased significantly with respect to Run 1. 

The first years of LHC operation have been a marked success, with the discovery of a new scalar particle, consistent with the Higgs boson~\cite{Khachatryan:2016vau}. Additionally, excellent agreement with the SM has been observed in precision measurements. 

\begin{figure}[h!]
\begin{center}
	\includegraphics[width=0.70\textwidth]{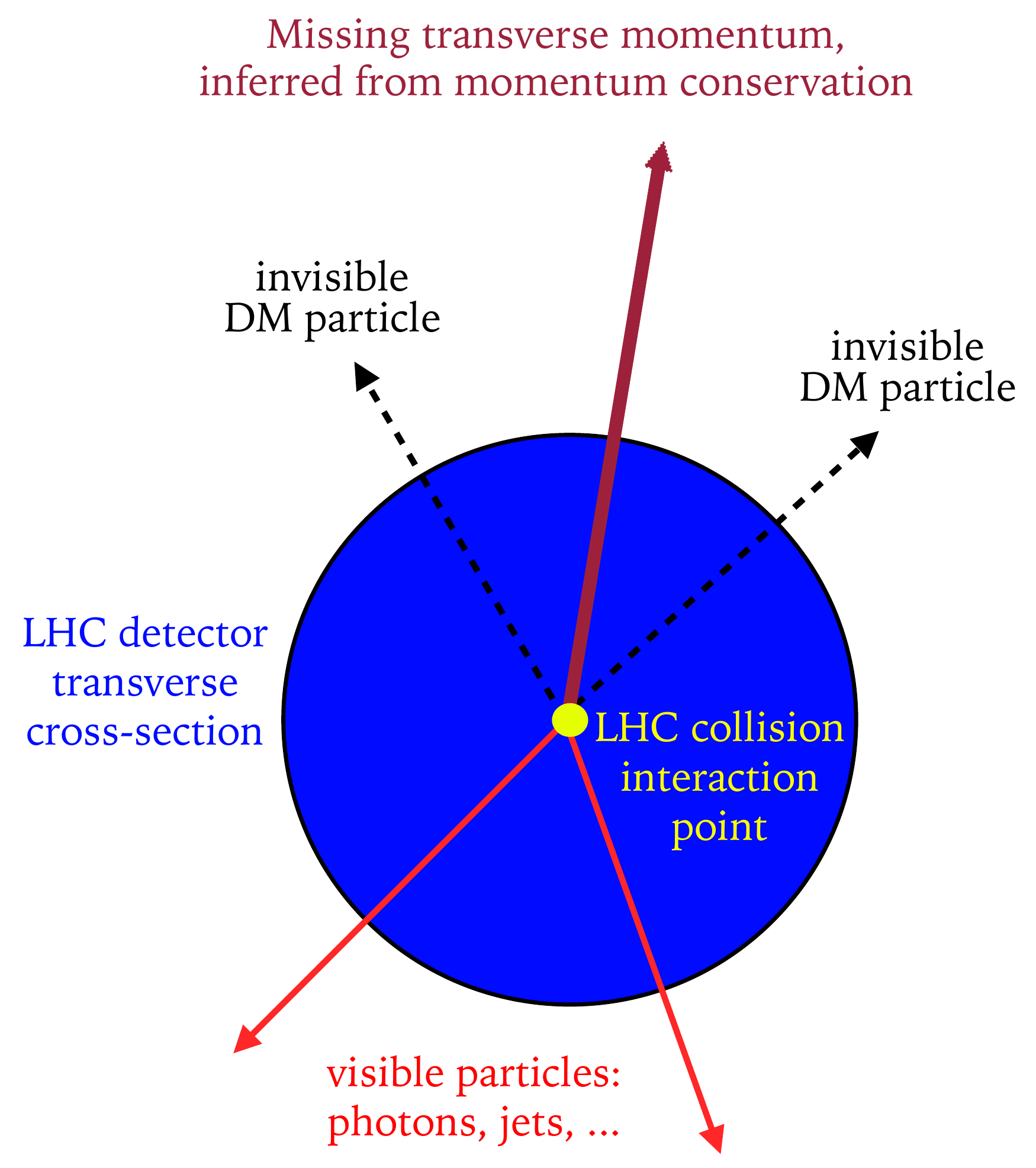}
\end{center}
  \caption{Schematic illustration of missing transverse momentum from DM production inferred from the recoil of visible particles, in a general purpose LHC detector.}
    	\label{fig:MET}
\end{figure}

DM particles will not produce a visible signal in collider experiments such as ATLAS and CMS, due to their extremely weak interaction with SM particles which constitute the detectors. The presence of DM particles can be inferred using transverse momentum conservation, as shown in the schematic drawing in Fig~\ref{fig:MET}.
The net momentum in the plane perpendicular to the colliding beams is zero, and must also be so after the collision has taken place. An imbalance in this plane, obtained as the negative vector sum of the transverse momenta of all detected particles, is the main signal for direct production of DM at colliders. This quantity is termed missing transverse momentum or missing transverse energy (\met).

\subsection{Predicting and observing DM at colliders}

Theoretical frameworks predicting WIMPs are an integral part of collider searches for DM. Predictions from the phenomenology of these models aid the design and optimization of experimental searches. In addition, it is only under the assumption of a specific theoretical framework that the connection between collider and non-collider searches can be made. Nevertheless, the wide range of possible properties of particle DM requires that the collider search program includes searches that are as model-independent as possible. 

Models of WIMP DM that would be observable at colliders vary in completeness and in complexity, in terms of number of parameters and specificity of the predicted phenomenology. 
Complete theories, such as SUSY, provide specific predictions for DM collider searches. Collider searches also target simpler benchmark models, modelling generic features of possible DM signals. 

\subsubsection{Searches targeting specific new physics models: Supersymmetry}

Questions surrounding the discovery and properties of the Higgs boson associated with the process of electroweak symmetry breaking motivate theoretical scenarios that predict new physics at the weak scale.
Many of these scenarios predict an invisible, stable heavy particle with mass close to the weak scale that can also serve as a DM candidate. Such scenarios predict many more new physics particles that can produce distinctive collider signals. 
Examples of this class of models include composite Higgs~\cite{Kaplan:1983fs}, new spatial dimensions~\cite{ArkaniHamed:2001nc,Agashe:2003zs,ArkaniHamed:1998rs,Randall:1999ee,Randall:1999vf} and SUSY (see Ref.~\cite{Chung:2003fi} for a review). 

SUSY has been a standard benchmark for collider searches for weak scale new physics and WIMP DM for many years, and it can be used as an example of a complete theory that includes a DM candidate.

\begin{figure}[h!]
\begin{center}
	\includegraphics[width=0.7\textwidth]{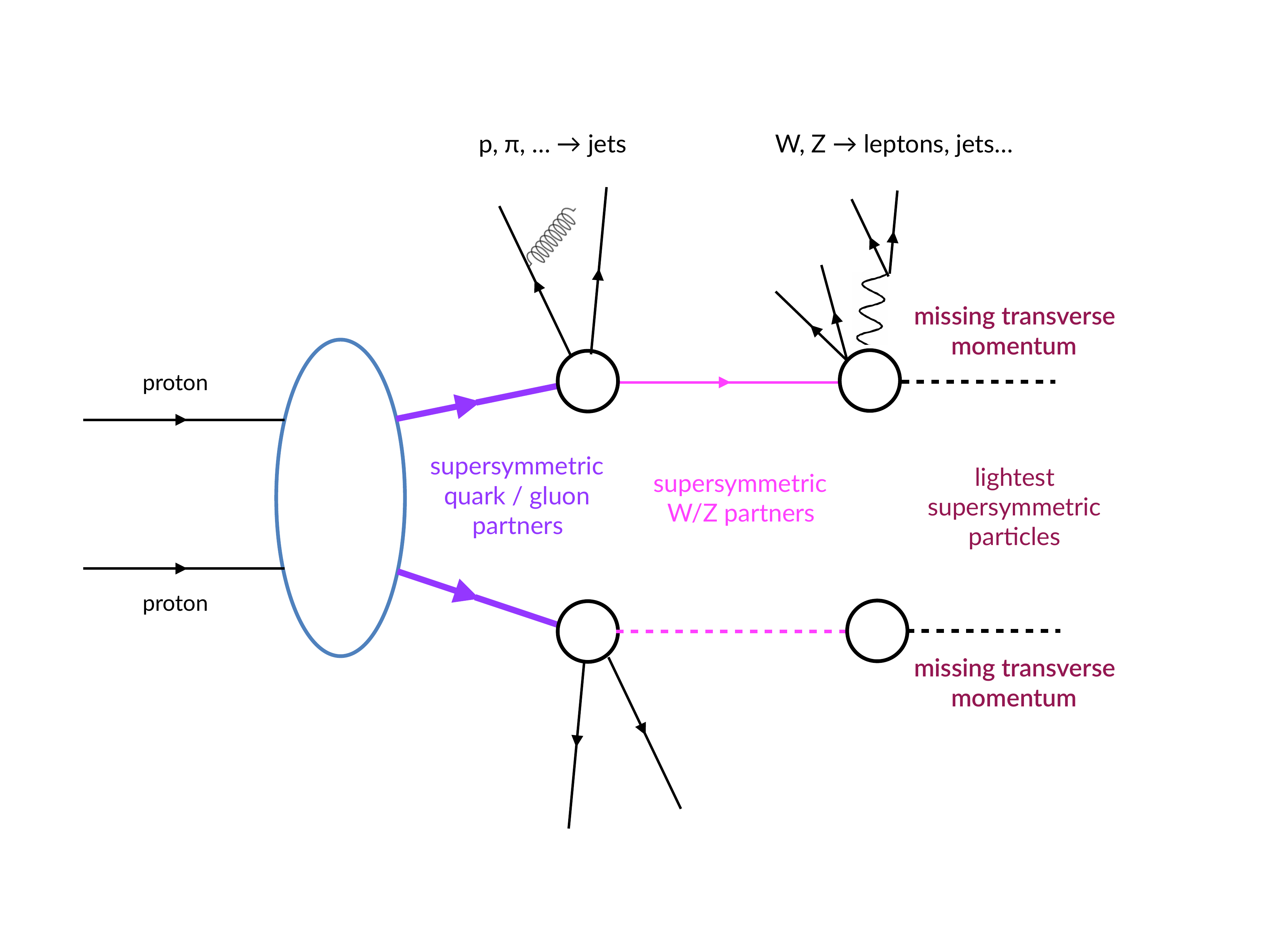}
\end{center}
  \caption{A typical production and decay chain involving SUSY particles. DM is at the end of the decay chain, leading to missing transverse momentum.}
  	\label{fig:SUSY}
\end{figure}

In SUSY models, each of the SM particles has a supersymmetric partner, with spin differing by a half-integer with respect to its counterpart. 
To comply with experimental observations where baryon and lepton number are conserved, a symmetry 
called \textit{R-parity} is often assumed in specific incarnations of SUSY models.
As a consequence of this symmetry,
the lightest SUSY particle (LSP) is stable and can be identified with the WIMP DM candidate.  Collider signals of SUSY are characterised by cascade decays of superpartner 
particles terminating in the LSP, as shown in Fig.~\ref{fig:SUSY}. 
These signals produce a final state signature in the detector that is rich
in collimated sprays of particles (jets) from quarks and gluons, in some cases accompanied by leptons and photons, alongside a significant amount of missing transverse momentum. 
Experimental search strategies for
SUSY searches discriminate signal and background events by the amount of missing transverse momentum, by the number of the other objects produced in the collision, and by  kinematic quantities that are functions of the visible particle masses and the missing transverse momentum in the event. 

A complete SUSY theory framework with full predictive power has a large number of parameters, as all superpartner masses and interactions with SM particles need to be specified. The benchmark models for many SUSY searches at the LHC are, rather than full theories, simplified SUSY models ~\cite{Alwall:2008ag}. Rather than predicting specific experimental signatures for one particular choice of model parameters, simplified models approximate the features of SUSY signals at LHC energies. 

Simplified models separate the masses of the SUSY particles involved in a limited number of decays from the rest of the superpartner particles, by  setting most of the SUSY particle masses above the collider center of mass energy. This simplifies the possible decay modes, fixes their relative rates and generally reduce the complexity of a complete SUSY theory while still describing the relevant phenomenology of LHC searches for individual SUSY particle production and decay modes. LHC results interpreted as simplified models can still be related to the more complex behaviour of SUSY theories (see e.g. Refs.~\cite{Kraml:2013mwa,Gutschow:2012pw}). 
So far, no experimental evidence for SUSY has been found. For details on the experimental constraints on SUSY particles see the experimental SUSY review in Ref.~\cite{Olive:2016xmw}.

\subsubsection{General searches for DM: effective field theories and simplified models}

\begin{figure}[h!]
\begin{center}
	\includegraphics[width=\textwidth]{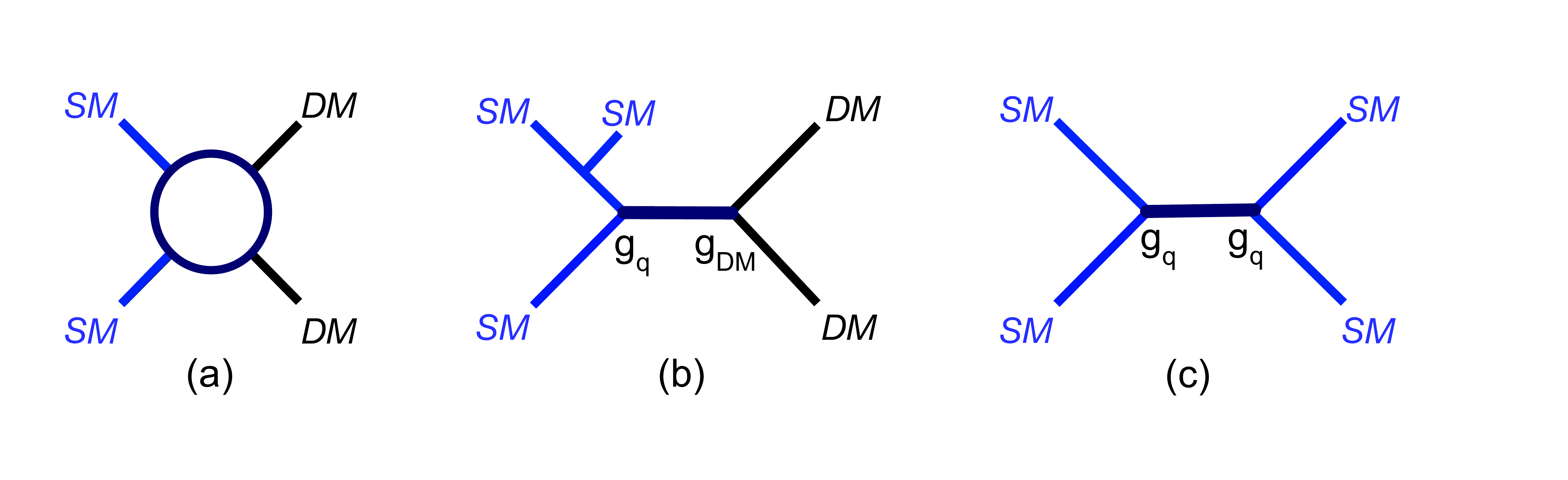}
\end{center}
  \caption{Schematic illustration of the basic SM-DM interactions at colliders, with time flowing from left to right. Sketch (a) shows the basic SM-DM interaction in an effective field theory (EFT), (b) shows its extension as a basic simplified model where a new mediator particle is exchanged in the $s-$channel (including an additional energetic object radiated from one of the initial state quarks) and (c) shows the same simplified model where the mediator decays back into SM quarks. The coupling constant characterizing the mediator-quark interaction strength is denoted as ($g_q$), while the mediator-DM coupling constant is denoted as $g_{DM}$. Adapted from \cite{MonoX}}
    	\label{fig:monox}
\end{figure}

While the searches in the context of complete theoretical frameworks such as SUSY remain a focus of the LHC, in recent years the DM search program has been augmented by a model--agnostic approach. 
The starting point for generic DM searches is the basic interaction between SM particles and WIMPs, as shown in Fig.~\ref{fig:monox} (a). The basic process sought at colliders is the pair production of DM, in association with one or more additional SM particles, as sketched in Fig.~\ref{fig:monox} (b). Events are selected for these types of searches if they contain a high momentum object (e.g. a jet, a photon, a vector boson...) radiated by the initial state quarks and gluons, in combination with significant missing transverse energy~(\MET). This signature is motivated by the assumption that proton-proton collisions produce two WIMPs, which remain undetected in the experiments but can be inferred from the measured \MET in the event as they recoil against the energetic mono-object. 
Given the presence of missing transverse momentum and of at least one highly energetic object, this search approach is sometimes called \textit{mono-X}~\cite{Birkedal:2004xn, Petriello:2008pu, Gershtein:2008bf, Cao:2009uw, Beltran:2010ww, Goodman:2010ku, Bai:2010hh, Fox:2011pm, Aaltonen:2008hh}. This has become a standard search strategy at colliders. 

Early LHC mono-X searches employed an~\textit{Effective Field Theory} (EFT) approach to interpret the experimental results in terms of DM~\cite{Goodman:2010ku} and compare them to results of DD and ID experiments. 
Within the EFT approach, SM and DM particles interact through a four-point effective (contact) interaction. This is akin to the Fermi theory describing weak interactions before the introduction of the W and Z bosons~\cite{Fermi2008}. 
Heavy particles mediating the SM-DM interactions are mapped to generic effective operators. These operators are characterized by the energy scale at which the SM-DM interactions would occur. 
They capture the characteristics of the SM-DM interaction as long as the states mediating the interaction are heavier than the operator's energy scale~\cite{DeSimone:2016fbz,Buchmueller:2013dya}. Searches at the LHC Run 1 constrained numerous operators up to energy scales of 1.5 TeV~\cite{Aad:2015zva, Khachatryan:2014rra}.

In its simplicity, the EFT approach is an attractive tool to encapsulate the relevant degrees of freedom and phenomenology for WIMP interactions without introducing the complexity of a full theory, but the EFT description of DM at colliders may break down at energies corresponding to the scales of the particles mediating the SM-DM interaction. 
It is therefore fruitful to take one step beyond the simple EFT approach in order to further probe the interaction between SM and WIMP DM. In a similar fashion as in SUSY models, but in this case in the direction of added complexity with respect to effective field theories, simplified models introduce particles that mediate the interactions between SM and WIMP DM. As in the case of the SUSY simplified models described earlier, these models describe the relevant LHC phenomenology and can be designed to be fully consistent at all energy scales. The phenomenology of simplified models where the mediator mass is large corresponds to that of an EFT. Typical simplified models used at the LHC only have a handful of parameters and allow the design and characterization of generic searches that need no assumptions about extended particle sectors. 
Simplified models represent a widely used class of benchmarks for the design and interpretation of DM searches at the LHC: Refs.~\cite{DeSimone:2016fbz,Abercrombie:2015wmb} (and references therein) provide reviews of the simplified models used for the LHC Run 1 and of their relation with EFTs. 

One of the simplest possibilities for the interactions described by simplified models is that the mediator particle is a SM particle, or has the same couplings as SM particles. If DM particles are part of a weak $SU(2)_L$  multiplet, they can interact with the SM through weak interactions mediated by the W and Z bosons. Scenarios with SM as mediator particles are constrained by electroweak precision measurements and direct searches~\cite{Escudero:2016gzx,
Cotta:2012nj}.
Other examples of related cases are supersymmetric theories where the Wino and the Higgsino are weak triplet and doublet, respectively. For these kinds of SUSY scenarios, much of the parameter space remains unexplored. The upcoming runs of the LHC can probe the WIMP DM mass in these SUSY models only up to a couple hundred GeV, requiring future colliders for further exploration \cite{Olive:2016xmw,Golling:2016gvc}.
Furthermore, the Higgs boson can also mediate SM-DM interactions. For example, Higgs portal scenarios are sought both via the invisible decays of the Higgs boson and via measurements of the Higgs couplings~\cite{Aad:2015pla, Khachatryan:2016whc}. The fraction of decays of the Higgs boson into invisible particles, including processes where the Higgs decays into DM particles, is currently measured to be less than 24\%. The sensitivity of those searches and measurements is still largely above the fraction of Higgs bosons decaying into four neutrinos via two $Z$ bosons predicted by the SM. 

Simplified models where the mediator particle is not yet included in the SM are used as the main benchmark models for LHC searches. An example is the case of a new mediator particle exchanged in the $s-$channel as in Fig.~\ref{fig:monox} (b-c). The SM-DM interaction is characterized by the mass of the new mediator particle, by the type and magnitude of the mediator's coupling strengths to the WIMP DM ($g_{DM}$) and to SM quarks ($g_q$), and by the DM particle mass. LHC searches interpreted in terms of this simplified model consider a new boson with either vector or axial vector couplings as the mediator particle, with the coupling to DM fixed to unity and coupling to quarks fixed at 0.25. 
These coupling strength values are free parameters of simplified DM models. In this case, they have been chosen so that they lie within the regime of validity for this kind of simplified model and so that strong constraints from direct mediator searches before the LHC Run 2 results are avoided~\cite{Chala:2015ama}.  

Within this simple benchmark scenario, mono-X searches at ATLAS and CMS constrain mediator masses up to 2 TeV and DM masses up to 500 GeV
~\cite{Aaboud:2016tnv,Aaboud:2016uro,Aaboud:2016qgg, Khachatryan:2016mdm}. Similar models include interactions with leptons, needed for the model's self-consistency  (see e.g. Ref.~\cite{Kahlhoefer:2015bea}), or include interactions between the new boson and the Higgs bosons (see e.g. Refs.~\cite{Carpenter:2013xra,Berlin:2014cfa}). 
The choice of a limited number of benchmark scenarios is still intended to represent a broader category of models where a Z-boson-like particle mediates the SM-DM interactions. The kinematic distributions of the signals to which collider and DD/ID searches are sensitive do not change significantly upon changing the mediator particle to a pure vector boson rather than an axial vector boson. However, modifying the mediator coupling type and values will affect the cross-section of this model in collider and DD/ID experiments differently. 

An alternative simplified model scenario to the exchange of a $Z$-boson-like particle is the one where SM-DM interactions are mediated by a new scalar (Higgs-like) or pseudoscalar particle, exchanged as in Fig.~\ref{fig:monox} (b-c). The scalar mediator's couplings to fermions are equal to the SM Higgs-fermion couplings, while the couplings to WIMP DM particles are set to unity. In more sophisticated versions of the same model, this mediator can mix with the Higgs boson (see e.g. Refs.~\cite{Carpenter:2013xra,Berlin:2014cfa}). 
LHC searches are starting to be sensitive to these benchmarks in mono-jet final states~\cite{Khachatryan:2016mdm} or in final states where the DM mediator is produced in association with heavy quarks. It is expected that with the 2016 dataset this category of simplified models will be explored in more detail.

Colliders also play a crucial role in the context of models with new particles mediating the interaction between WIMP DM and SM particles, as they can search for the visible decays of the mediator particles.  One guaranteed visible mediator decay at colliders is the decay into the same SM particles whose collisions produced these mediators, namely quarks and gluons. 
Although such a signal would not involve the dark matter particle directly, it is an integral part of the model and thus also probes the dark matter hypothesis. Heavy ($\sim$ TeV) resonant states decaying into a pair of quarks or gluons would produce a distinctive signal at hadron colliders: a narrow excess in the invariant mass of the two most energetic jets (dijet) atop the smoothly falling QCD background. Searches for these new particles are only loosely tied to specific models (see Ref.\cite{Harris:2011bh} for a review and list of benchmark signals). Even though the connection between a signal in one of these searches and Dark Matter needs confirmation from direct searches for DM particles, these searches are sensitive to the presence of DM mediators with couplings to quarks and gluons. The invariant mass of the two most energetic jets in the event (di-jet system) is used as the main observable for the search.
Mediator particles can also have couplings to other SM particles and therefore can decay to other final states. We will restrict the discussion in this article to the example of a $Z$-like mediator that interacts only with quarks and DM particles with axial vector couplings.

Searches for DM mediators with masses above the TeV scale~\cite{ATLAS:2015nsi, Sirunyan:2016iap} benefit from the combination of the higher Run 2 collision energy and data rates: their mass reach will be approximately doubled in this LHC data taking period. With their high-mass searches, ATLAS and CMS dijet searches constrain axial vector mediator masses from 1.1 TeV to 2.5 TeV in the axial vector scenario used for mono-X searches above. As the parameter space for the mass of the mediator particle in these simplified DM models is only loosely constrained by the relic density~\cite{Fairbairn:2016iuf,Chala:2015ama}, searches for particles with mass above the TeV are not sufficient to cover the full parameter space. This motivates additional interest in rare, low-mass resonances not yet excluded by previous collider searches~\cite{Harris:2011bh}, which benefit from the higher collision rates. Whilst there is no issue in recording all background and possible signal events at the highest invariant masses, only a reduced fraction of events from high-rate processes at lower invariant masses can be fully recorded during data taking. To overcome this limitation, LHC searches for particles with invariant masses below 1 TeV either only record a limited amount of information rather than the full event~\cite{Khachatryan:2016ecr}, or target new particles that are produced in association with a highly energetic jet or from radiation of the initial state quarks and gluons (ISR). One example of such a final state is given by the process in Fig.~\ref{fig:monox} (b), with the difference that the final state DM particles are replaced by SM quarks~\cite{An:2012va}. These searches can probe mediator masses as low as 100 GeV.

\begin{figure}[h!]
\begin{center}
\centering
	{\includegraphics[width=0.9\textwidth]{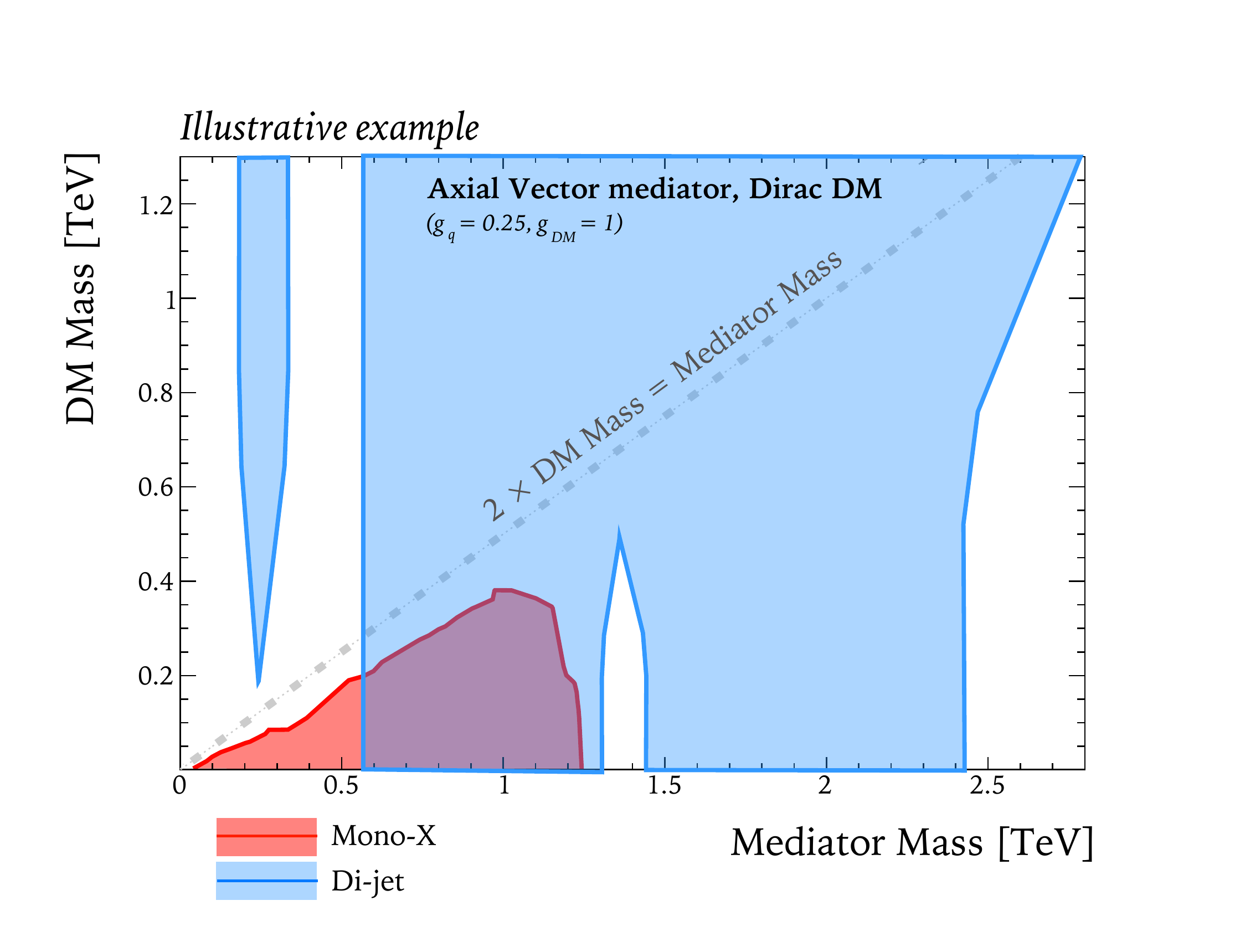}}
\end{center}
  \caption{Sketch of the constraints on a simplified model of WIMP DM where a particle with axial vector couplings of 1.0 and 0.25 to DM and SM respectively is exchanged. The constraints from mono-X (exclusion region in red) and dijet DM searches (exclusion region in blue) are shown in the plane of dark matter mass vs. mediator mass. The couplings represented in this sketch and its general features are inspired by LHC dijet and mono-X results \cite{Sirunyan:2016iap, Sirunyan:2017onm,ATLAS:2015nsi,Aaboud:2016tnv}.}
    	\label{fig:dijetDM}
\end{figure}

The complementarity between searches for WIMP DM particles and searches for mediator particles using the axial vector simplified model is illustrated in the sketch of constraints from mono-X and dijet searches in the DM vs mediator mass plane, in Fig.~\ref{fig:dijetDM}. 
Mono-X searches cover the model parameter space where the mass of the DM particle is lower than two times the mediator mass, allowing the mediator to decay on-shell to DM particles. Dijet searches are more sensitive at higher mediator and higher DM masses with respect to mono-X searches, as in that region of the parameter space DM decays of the mediator are suppressed with respect to dijet decays. The parameter space covered by mono-X and dijet searches is however highly dependent on the choice of quark couplings.

\section{Common interpretation of collider and non-collider results}
 
The comparison of collider results with non-collider results involves a common theoretical model, therefore requiring additional assumptions to be taken into account. For example, for non-collider results these assumptions include the knowledge of relic density and the DM density in the vicinity of the earth or of the galactic center, while for collider results details of the production and decay mechanisms in particle collisions have to be specified. 
These comparisons are often performed using observables used for the presentation of results of non-collider searches, namely the cross-section of the annihilation of DM particles to SM particles or the cross section of the scattering between nucleons and DM particles.

The parameter space of complete new physics models such as SUSY is inherently linked to the constraints on WIMP DM
(see for example~\cite{Aaboud:2016wna,Aad:2015baa,Khachatryan:2016nvf,Bagnaschi:2015eha,Baer:2016ucr}). SUSY searches at colliders and searches from DD/ID experiments, as well as flavour physics and precision electroweak measurements and the measured value of the DM relic density, can be used to constrain the parameter space of full SUSY models, assuming that the LSP is the DM candidate. 

\begin{figure}[h!]
\begin{center}
\centering
	\subfloat[] {\includegraphics[width=0.7\textwidth]{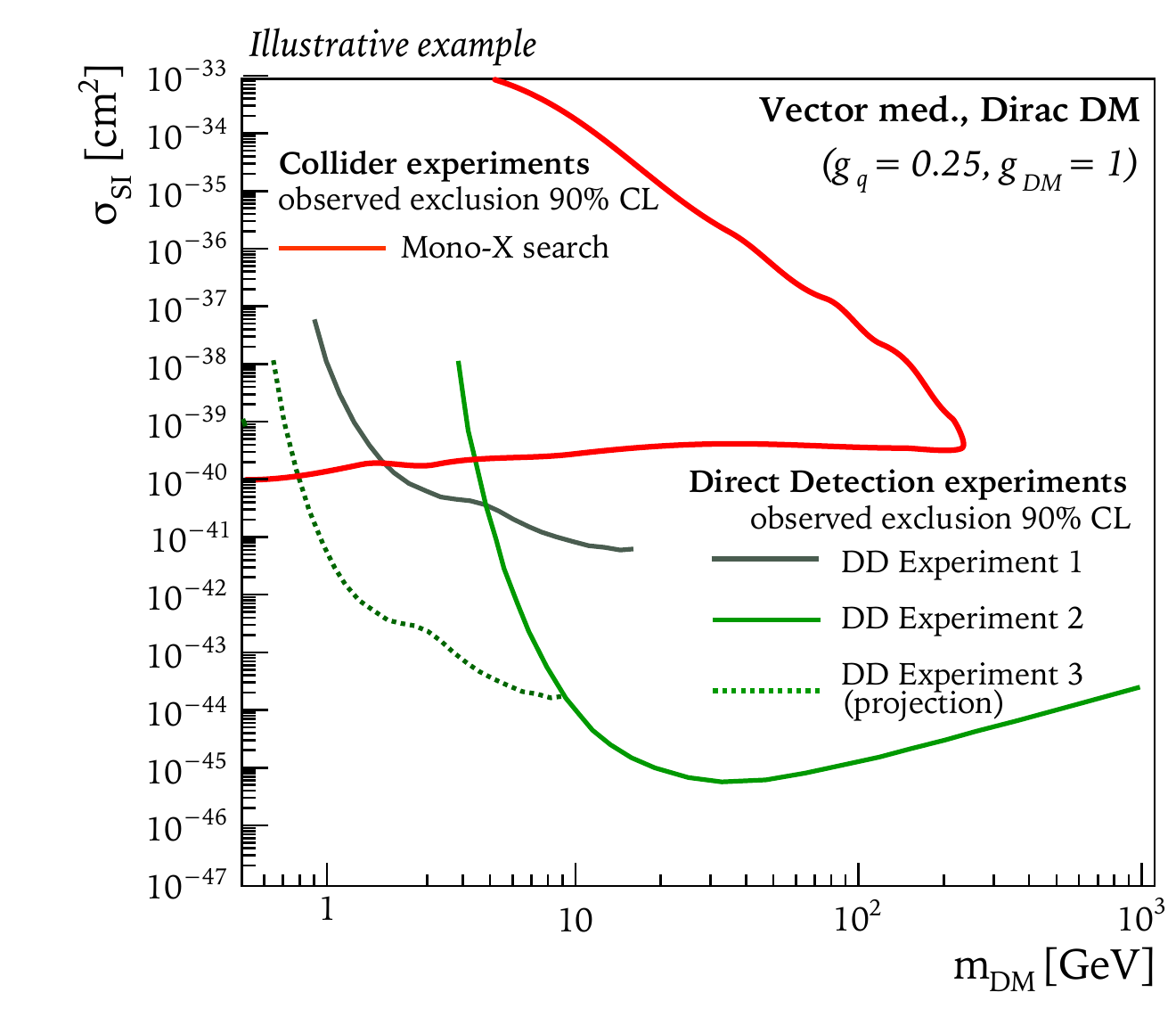}}
\end{center}
  \caption{
  Sketch of the comparison between constraints set on spin-dependent WIMP--proton scattering cross section for DM searches at colliders (red line) and at direct detection experiment probing different regions in WIMP mass (green lines). The model parameters represented in this sketch and its general features are inspired by the results of LHC monojet and DD searches \cite{Sirunyan:2017onm,Aaboud:2016tnv,Agnese:2016cpb,Angloher:2015ewa,Akerib:2016vxi}, but this figure only has an illustrative purpose. This comparison is performed within a simplified model in which a vector boson is exchanged to mediate the SM-DM interaction with DM and SM quark couplings fixed to unity and 0.25 respectively. It should be noted that the the absolute exclusion of the collider result in this kind of comparisons, as well as its relative importance, will depend on the chosen coupling and model scenario. Therefore, the comparison of collider limit with DD excluded regions in this kind of plots is not applicable to other choices of coupling values or models.
}\label{fig:DMComparison}
\end{figure}

As described in detail in~\cite{Boveia:2016mrp}, a collider and non-collider searches can also be compared when interpreted in the framework of simplified models. In this approach, constraints on simplified models from collider searches are translated into upper limits on the DM-nucleon scattering cross section, and analyzed as a function of the mass of the DM particle.
The details of these comparisons depend on the assumed interaction structure and coupling scenarios of the simplified model. As an illustrative example, Ref.~\cite{Boveia:2016mrp} recommends an example coupling scenario to compare the results of collider and DD experiments. A sketch of such a comparison is shown in Fig.~\ref{fig:DMComparison}. We note that this is just one example, and the details of such comparisons depend on the model assumptions. At the same time,  such comparisons in the framework of simplified models show that the sensitivity of collider experiments can complement that of non-collider experiments for different interaction structures and certain kinematic regions. For example, this could be the case for low DM masses where the direct detection is limited by experimental thresholds.    

We also note that there have been  recent effort of non-collider experiments towards improved sensitivity at low DM masses, see e.g.~\cite{PhysRevLett.116.071301, Angloher:2015ewa,Strauss2016}.

\section{Conclusions and outlook}

As the LHC has just concluded the delivery of only about 1\% of the dataset planned during its running period up to 2035, it is evident that the search for WIMP DM at collider experiments has only just begun. 
DM at colliders offers a rich spectrum of possible signals. While DM can be part of a large framework of TeV-scale new physics with striking signals, such as SUSY, it can also be represented within minimal frameworks that only allow for simple interactions with the SM. As summarized in this article, LHC searches have covered both ends of the range of theoretical models, and continue to make progress. DM searches at colliders play an important role in complementing the search programme of non-collider searches.

In any case, given our limited knowledge about DM, it is essential to cast a wide net and broaden the scope of collider searches beyond the minimal WIMP hypothesis. There are many possible directions, some of which are already pursued by collider searches. For example, it is possible that DM is part of a larger dark sector with a rich structure rather than a single particle~\cite{Alexander:2016aln}. This dark sector can include additional particles, which can couple to the DM and have interesting dynamics. DM could have self interactions~\cite{Spergel:1999mh}, or there may be more than one species of DM. A typical feature of dark sector models alternative to WIMP DM is the presence of many more characteristic signatures, complementing the missing transverse momentum from invisible particles. More often than not, the most prominent signal does not stem from the observation of the DM particle itself. Therefore, it is also fruitful to consider collider searches for dark sectors in a broader class of experimental signatures. LHC searches for dark photons and long-lived particles, for example, are a step in this direction~\cite{Ilten:2016tkc,Curtin:2014cca,Aad:2015uaa,Aad:2014yea,Khachatryan:2016sfv,CMS:2014wda}. Dark sector models also offer signals beyond the standard DM detection methods, involving cosmological observations and other terrestrial experiments such as fixed target searches, providing a rich unexplored experimental landscape in the search for DM. 

The collider searches described in this review offer an essential contribution to probing the WIMP dark matter hypothesis and complement the searches at DD and ID experiments: a discovery would be revolutionary. However, the search for WIMP dark matter at colliders could also be a long journey which may not end with the LHC, but be continued at future, higher energy colliders.

\clearpage

\begin{center}

\fbox{\begin{minipage}{\textwidth}

\section*{Box 1 | The WIMP ``miracle"}
\footnotesize
DM is only observed experimentally through its gravitational interaction. There is little knowledge about the identity and cosmological evolution of DM particles. Among the vast number of possible scenarios for the nature of DM, Weakly Interacting Massive Particles (WIMP) stand out as a compelling theoretical possibility. This scenario starts from the simple assumption that DM particles have small but appreciable interactions with SM particles, so that they are in thermal equilibrium in the early universe. In addition to giving rise to a simple initial condition for the cosmological evolution of the DM abundance, this assumption also allows us to make experimental predictions involving DM particles without needing to know their earlier cosmological evolution. DM particles could still have a large range of masses and couplings with the SM particles. However, if the DM mass is around the weak scale (similar to the massive particles found in the SM) and if DM particles couple with SM particles with a strength similar to that of the SM electroweak interactions, this simple scenario approximately predicts the correct observed DM relic abundance. This concordance is sometimes called WIMP ``miracle", even though it is more of a ball-park estimate than a precise prediction. If such WIMP DM was detected in a laboratory and its properties measured, we could then precisely predict the WIMP relic abundance and compare this prediction with cosmological observations. The LHC and other experiments can probe a wide region in the mass-coupling space of WIMP DM. Even though the scenario of WIMP DM is by no means the only one, it is a simple and compelling scenario, and testing it thoroughly is a necessary milestone in our quest for the DM. 
\end{minipage}}

\end{center}

\begin{appendix}

Liantao Wang is the corresponding author for this manuscript, to whom correspondence and requests for materials should be addressed.
We thank A. Boveia, M. Danninger, A. De Roeck, T. J. Khoo, G. Landsberg, and C. Young for valuable comments on this manuscript.
Work by CD is part of a project that has received funding from the European Research Council (ERC) under the European Union's Horizon 2020 research and innovation programme (Grant Agreement No 679305) and from the Swedish Research Council. The work of LTW was supported in part by the Kavli Institute for Cosmological Physics at the University of Chicago through grant NSF PHY-1125897 and an endowment from the Kavli Foundation and its founder Fred Kavli.\end{appendix}

\bibliographystyle{naturemag}
\bibliography{nature_physics_DM}

\end{document}